\documentclass[twocolumn,showpacs,preprintnumbers,amsmath,amssymb,prl]{revtex4}
\usepackage{times}
\usepackage{graphicx}
\usepackage{amsmath, amsthm, amssymb}
\usepackage{latexsym}
\usepackage{amssymb}
\usepackage{bm}
\usepackage{dcolumn}
\usepackage{epstopdf}

\begin{document}
\title{Spin-Density Wave near the Vortex Cores of Bi$_2$Sr$_2$CaCu$_2$O$_{8+\delta}$}

\author{A.M. Mounce,  S. Oh, S. Mukhopadhyay, and W.P. Halperin*}
\affiliation{Department of Physics and Astronomy, \\Northwestern  University, Evanston, IL 60208,
USA}
\author{A.P. Reyes,  and P.L. Kuhns}
\affiliation{National High Magnetic Field Laboratory, Tallahassee, FL 32310,
USA}
\author{K. Fujita, M. Ishikado, and S. Uchida}
\affiliation{Department of Physics, \\University of Tokyo, Tokyo 113-8656, Japan}
\date{Version \today}


\begin{abstract} Competition with magnetism is at the heart of high temperature superconductivity, most intensely felt near a vortex core. To investigate vortex magnetism we have developed a spatially resolved probe using nuclear magnetic resonance.  Our spin-lattice-relaxation spectroscopy is spatially resolved both within a conduction plane as well as from one plane to another.  With this approach we have found a  spin-density wave associated with the vortex core in Bi$_2$Sr$_2$CaCu$_2$O$_{8+\delta}$, which is expected from scanning tunneling microscope observations of "checkerboard" patterns in the local density of electronic states.\cite{hof02}   We determine both the spin-modulation amplitude and decay length from the vortex core in fields up to $H=30$ T. 

\end{abstract}

\maketitle

\newcommand{\dg}{^{\circ}}

\DeclareGraphicsRule{.tif}{png}{.png}{`convert #1 `basename #1 .tif`.png}
\normalsize
\thispagestyle{empty}

Superconductivity and magnetism have been in the forefront of the study of high temperature superconductivity, largely because their interplay is central to the pairing interaction from which superconductivity originates.  An important manifestation of this dichotomy can be found near a vortex core where circulating supercurrents create a sufficiently large magnetic field as to suppress the superconducting state, thereby tipping the balance for stability in favor of magnetism.\cite{aro97}  A number of spin-sensitive, spatially-resolved, probes  have been employed to explore this vortex state, including muon spin resonance, small angle neutron scattering, and nuclear magnetic resonance (NMR).  These experiments can give the spatial distribution of local magnetic fields, as in the example of the  $^{17}$O NMR  spectrum shown in Fig.~1.  This spectrum, as well as results of other methods, are a superposition of local fields from both magnetism and superconductivity and their deconvolution can be quite difficult.   \\
\begin{figure*}[t]
	\centering
	\includegraphics{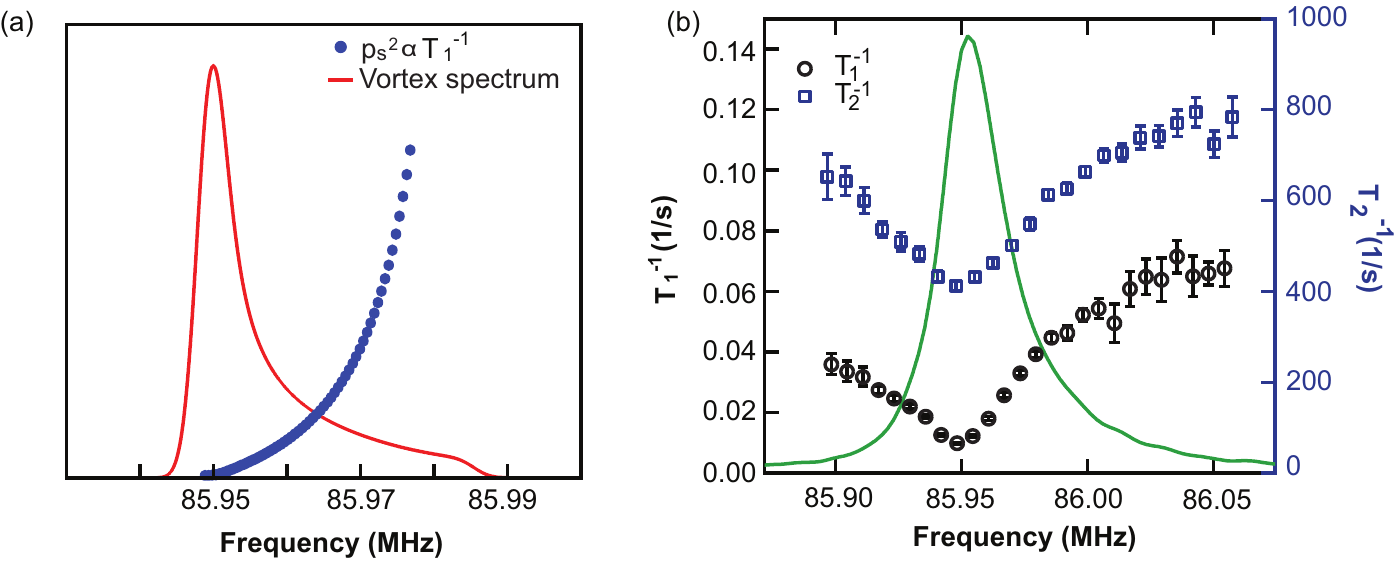}
	\caption{The local field distribution and relaxation rate distribution in the vortex unit cell at $H=14.9$ T, (a) calculated from Ginzburg-Landau (GL) theory.\cite{bra97} The super current momentum squared is proportional to the Doppler term, and is the dominant contribution, in the spin-lattice relaxation rate. (b) The relaxation rates across the $^{17}$O NMR  spectrum at $T=4.2$ K of a slightly overdoped, $T_c=82$ K  Bi2212 single crystal.  On the low frequency side of the peak in the spectrum there is an unusual and  substantial increase in the spin-lattice relaxation rate, $T_1^{-1}$, mirrored in the spin-spin relaxation rate, $T_2^{-1}$.}
	\label{T1andT2}
\end{figure*}
\indent Here, we develop a new approach based on  NMR spin-lattice relaxation (SLR) rate spectroscopy.  The source for relaxation is the Doppler effect on the nodal quasiparticle excitations and is given by the square of the supercurrent momentum, $p_s^2$.  Consequently, SLR can be attributed solely to superconductivity.  Being two-dimensional, the rate is fully resolved from one superconducting plane to another, a decided benefit in comparison with measurements of the local field which average in three dimensions.  Most importantly, SLR has strong spatial contrast increasing with the inverse square distance from the vortex core, $\sim r^{-2}$.  Taking advantage of this spectroscopy, together with measurements of the NMR spectrum of local magnetic fields, we are able to characterize the magnetic spin density wave (SDW) associated with vortices in our sample, consisting of a slightly over-doped, single crystal of the anisotropic high temperature superconductor, Bi$_2$Sr$_2$CaCu$_2$O$_{8+\delta}$ (Bi2212),  $T_c=82$ K.  Our measurements extend over a wide range of applied magnetic field, 4 T $< H \leq$ 30 T. \\
\indent Scanning tunneling microscopy (STM) experiments on Bi2212 at moderate magnetic fields and low temperatures showed that vortices induce a checkerboard pattern of the local density of states with a spatial period of 15 \AA \,$\approx$ 4$a_0$.\cite{hof02}  Further STM studies found such electronic ordering in Bi2201\cite{han04} and NaCCOC\cite{wis08} persisting to non-superconducting dopings and above the superconducting transition temperature. Elastic neutron scattering experiments\cite{lak01, lak02, kha02} on the cuprate superconductor, La214, produced magnetic Bragg reflections corresponding to a periodicity of 30 \AA  $\approx$ 8$a_0$.  Although spatially unresolved, this result was associated with the checkerboard pattern found from STM on Bi2212, since the period of a spin density wave is expected to be twice that found in the density of states.\\
\begin{figure}[b]
	\centering
	\includegraphics[width=.45\textwidth]{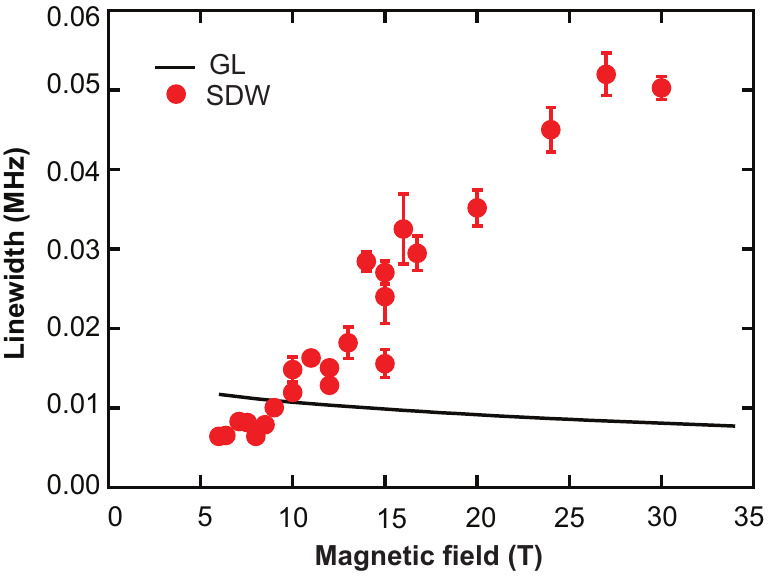}
	\caption{The NMR linewidth as a function of applied magnetic field at 4.2 K.  The contribution to the local field distribution from supercurrents was calculated from GL theory (solid curve) and was subtracted from the data points.}
	\label{AveT1}
\end{figure}
\indent There have been many theoretical interpretations for these results including competing SDW and superconducting order\cite{dem01,dem02}, a Wigner crystal of Cooper pairs/pair density wave\cite{che04, tes04}, striped ordering\cite{kiv03}, and $d$-density wave checkerboard order\cite{seo07}. More recently, a phenomenological octet model of quasiparticle scattering from the fermi arcs\cite{mce03} has gained popularity in interpreting STM and STS results.\cite{han09} The octet model fits some STM, STS, and ARPES\cite{dam03} results, however, this model does not account for the neutron scattering results\cite{lak02} as well as the STM results of NaCCOC\cite{wis08}.  It has been suggested\cite{sac02} that, by increasing the applied magnetic field, magnetic order might be "tuned" and further suppress superconducting order.  Since NMR is a bulk, spin-sensitive probe and can be performed up to high field, it is ideal for investigation of the vortex states of high temperature superconductors.  In the present work, we exploit these capabilities of NMR in addition to its potential for spatial resolution. \\
\indent In the case of an ideal, non-magnetic, superconductor with straight vortex lines that form a two-dimensional lattice, the lowest frequencies of the NMR spectrum, Fig.~1(a,b), come from the resonant nuclei in the sample positioned furthest from the vortex cores; the highest frequency  components are from the nuclei closest to the vortex cores  where the field is highest; and the peak  corresponds to a saddle-point in the distribution of frequencies $\omega =\gamma H$, where $\gamma$ is the nuclear gyromagnetic ratio.  This relation between spatial positions in the vortex lattice unit cell and the NMR frequency is well-established.\cite{bra91}  However, contributions to the local fields from magnetism associated with vortex cores must be accounted for independently.\\
\begin{figure*}
	\centering
	\includegraphics{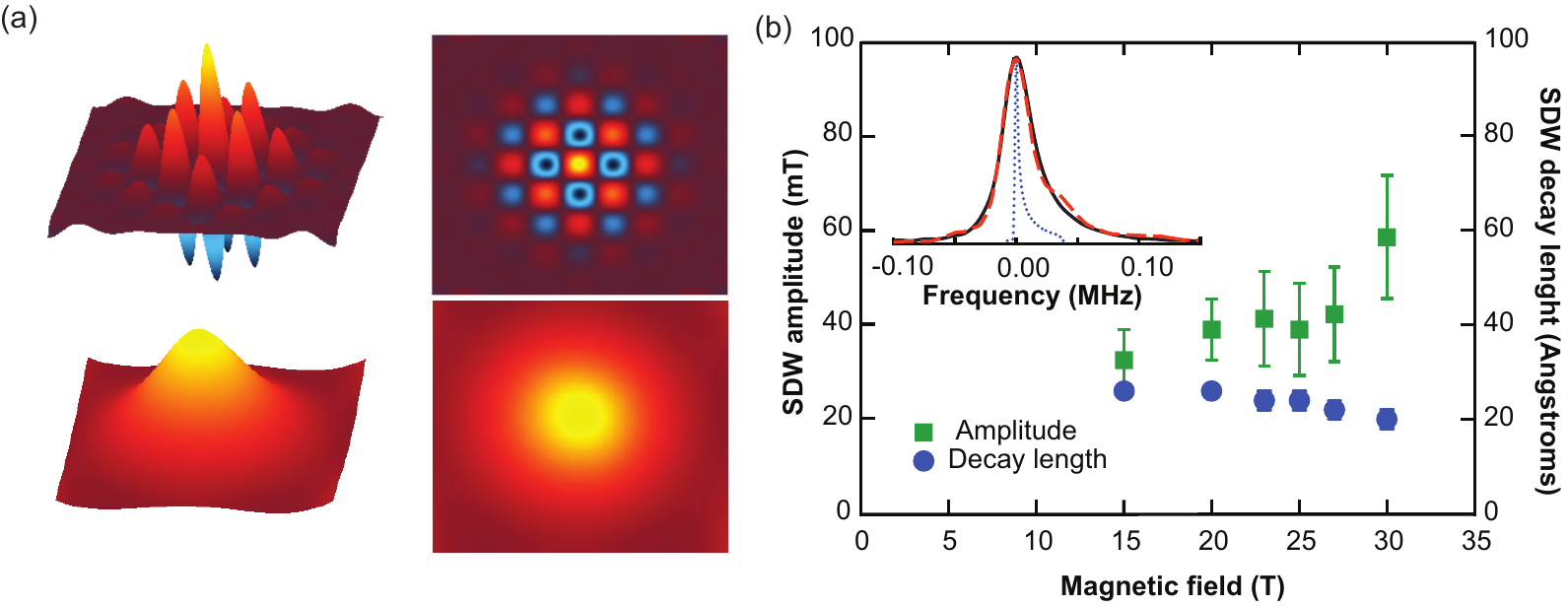}
	\caption{(a) The spatial distribution of the local magnetic fields near a vortex within a CuO$_2$ plane calculated for: (top) a spin density wave (SDW), Eq.~1, and (bottom) vortex supercurrents from GL theory\cite{bra97} for which the amplitude $A_{dia}$ has been increased by a factor of 3 with respect to $A_{SDW}$ for clarity. (b)The amplitude $A_{SDW}$ (squares) and decay length $\tau$ (circles) for the best fit spectrum generated from the local magnetic fields in (a).  There is an increase in $A_{SDW}$ and $\tau$  nearly constant with increasing magnetic field.  [Inset]  A comparison of the experimental (black), GL solution (blue, dotted), and SDW+GL (dashed-red) spectra centered at the central peaks, $H=15$ T.} 
	\label{ATvsH}
\end{figure*}
\indent A clear indication that such contributions exist in Bi2212 is evident from the spectrum of  the central transition for the O(1) planar site of $^{17}$O NMR, Fig.~1(b) and in the linewidths, Fig.~2.  We find the distribution of local magnetic fields to be much greater than that expected for vortex supercurrents alone.  Moreover, the linewidth increases systematically with applied magnetic field, rather than decreasing  as would have been the case for supercurrents represented by the solid curve in Fig.~2.  Secondly, in Fig.~1(b) and Fig.~4,  the SLR rate,  $T_1^{-1}$, is strongly non-monotonic across the spectrum contrary to predictions\cite{mor00} for $d$-wave superconductors without magnetic vortices, Fig.~1(a).  Since SLR is from Doppler-shifted, nodal quasiparticles,  it must increase smoothly and monotonically with decreasing distance approaching the vortex core (see supporting online materials).  Thirdly, it is expected that the dominant interaction for spin-spin relaxation is from vortex vibrations which are also determined by vortex supercurrents.\cite{lu06}  Although there is not yet a quantitative theory for spin-spin relaxation,  nonetheless it should mimic SLR.  From Fig.~1(b), the strong similarity between the two rate profiles indicates that this is correct.  As with SLR, the spin-spin relaxation rate is not a monotonic function of local field.  \\
\indent From these results we infer that the local field distribution is not that of an ideal superconductor and has substantial magnetic contributions.  Furthermore, the well-defined correlation between NMR relaxation and position in the spectrum requires that these additional contributions be closely associated with vortices.  In fact, the non-monotonic behavior of the relaxation rate profiles can only be understood if the vortex local fields are oscillatory.  We find that a spin-density wave model for vortex magnetism can completely describe both field distribution (spectrum) and supercurrent distribution (SLR) and thereby characterize the vortex, spin-density wave state in Bi2212. \\
\indent In order to model the measured local field distributions, we construct a spin "checkerboard" cosine wave decaying as a gaussian away from the vortex core.  Consistent with STM and elastic neutron scattering measurements\cite{hof02,lak02} we take the period to be $\lambda = 30 $ \AA $ \sim 8a_0$ and allow the amplitude and range of the spin-density modulation to be free parameters for fitting to the  NMR spectra,
\begin{equation}
	B(x,y)= \ A_{SDW}  cos(2 \pi x/\lambda)cos(2 \pi y/\lambda)e^{(-(x^2+y^2)/2\tau^2)}
	\label{para}
\end{equation}	
\noindent where $A_{SDW}$ is the amplitude of the magnetization at the vortex core and $\tau$ is its decay length.  Eq.~1 produces a SDW magnetic field component as shown in the upper part of Fig.~3(a).\\
 \begin{figure*}[t]
	\centering
	\includegraphics{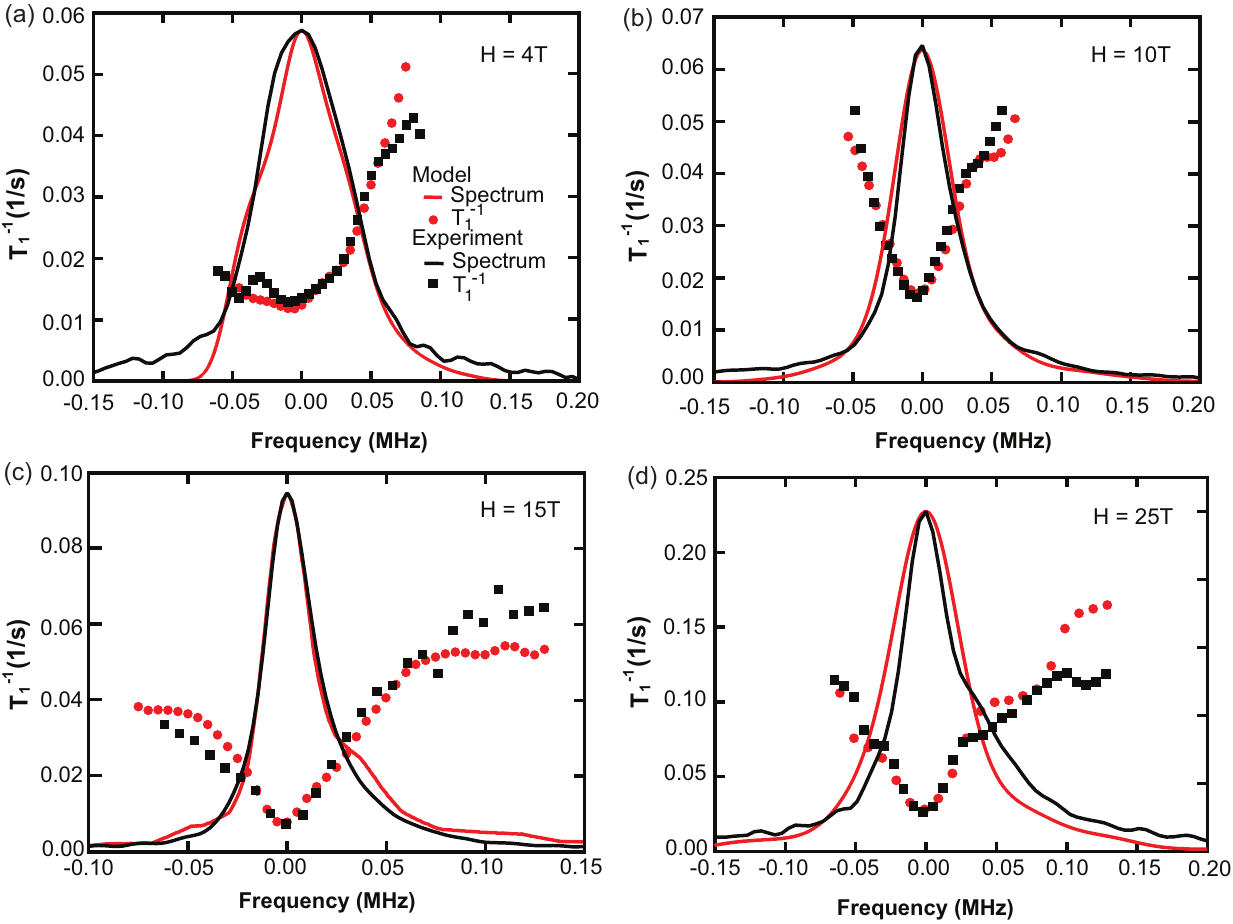}
	\caption{The calculated (red) and experimental (black) results for the spectra (curves) and relaxation rates (dots, model; squares, experiment) for various $H$ at 4.2 K,  (a-d) for ($H = 4, 10, 15, 25$ T). At low magnetic fields there is a near monotonic relationship for $T_1^{-1}$ as a function of frequency (local field) as expected for the case $A_{SDW}\leq A_{dia}$, see Fig.~1(a). At high magnetic fields the SDW dominates producing a non-monotonic distribution of relaxation rates.  It is clear that structure at high frequency in the experimental rate is replicated in the model calculation.  This structure is sensitive to the spin wave period (see supporting online materials).  The spectra are normalized to peak height and the model calculation for  $T_1^{-1}$ is offset vertically and scaled to compare with experiment.}
	\label{fields}
\end{figure*}
\indent We also include in our fitting the  local magnetic field distribution from diamagnetism of the vortex supercurrents and we add this contribution to Eq.~1 at each position in the vortex unit cell before fitting to the data.  The supercurrent diamagnetism was calculated using Brandt's algorithm obtained by solving the Ginzburg-Landau (GL) equations for an ideal vortex lattice\cite{bra97} and is shown at 15 T in Fig.~3(a) for a quantitative comparison with the corresponding SDW.  To find the proper fitting parameters, we calculated the spectra for the GL solution plus SDW for many different values of $A_{SDW}$ and $\tau$ and we used a chi-squared test to find the parameters shown in Fig.~3(b).  With increasing applied field $H$ there is an increase in $A_{SDW}$ and a nearly constant  $\tau$, approximately twice the superconducting coherence length, $\tau \approx 2\xi_o \approx 22$ \AA.   We then calculated the SLR rate being proportional to $p_s^2$, plus a constant offset, at each applied magnetic field where the supercurrent momentum is also calculated from GL theory.   Multiple values of the calculated SLR rate, within each of 32 contiguous intervals of local field, were averaged and then presented as a single value to be compared with the data in Fig.~4.  Over the entire range of applied field  we capture an excellent representation of both the field distribution and the relaxation profiles with just two parameters, despite the unusual non-monotonic behavior of the SLR rate as a function of frequency (local field).\\
\addtocounter{figure}{-4}
\renewcommand{\thefigure}{S\arabic{figure}}
\indent There are three ranges of applied magnetic field which characterize different profiles of $T_1^{-1}$ across the NMR spectrum, depending on the relative contributions of the diamagnetic local fields $A_{dia}$ and the vortex-induced magnetic contribution $A_{SDW}$.  At low magnetic fields, 4-6 T, when $A_{SDW} < A_{dia}$, $T_1^{-1}$ resembles the near monotonic behavior expected for a $d$-wave superconductor without magnetic vortices,\cite{mor00} see Fig.~1(a). Upon increasing the external magnetic field to 6-12 T there is a sharp increase of $T_1^{-1}$ on the low frequency side of the spectrum in the range where $A_{SDW} \gtrsim A_{dia}$.  Finally, at high magnetic fields, 15-30 T, there is a significant increase of $T_1^{-1}$ on both low and high frequency sides of the spectrum corresponding to $A_{SDW} >> A_{dia}$.   Measurements in YBCO have revealed a small upturn in $T_1^{-1}$ at low frequency, but much less pronounced than what we report here for Bi2212.  In the YBCO work, contamination on the low frequency side of the spectrum from quadrupolar transitions precluded an unobstructed view of the magnetic field distributions over the same wide range as is possible with Bi2212.  We think it is possible that vortices in YBCO are accompanied by a SDW checkerboard pattern which has not been resolved in these earlier NMR experiments.	\\
\indent In summary we have shown that a spin-density wave can be associated with vortices in Bi2212 which accounts for both the NMR spectrum and NMR relaxation behavior as a function of local magnetic field.  The SDW amplitude increases with applied field and its spatial range is nearly constant at $\approx 2 \xi_o$ up to $H=30$ T.  These results establish a link between previously reported checkerboard STM patterns\cite{hof02} and spin-dependent elastic neutron scattering in a cuprate superconductor\cite{lak02} and provide a detailed characterization of magnetic vortices in Bi2212. 

\section{ACKNOWLEDGEMENTS}	
	We thank H. Alloul, C.A. Collett,  M.R. Eskildson, W.J. Gannon, J.E. Hoffman, J.B. Ketterson, Jia Li, K. Machida, V.F. Mitrovi\'c, D.K. Morr, J. Pollanen, J.A. Sauls, Z. Te\ifmmode \check{s}\else \v{s}\fi{}anovi\ifmmode \acute{c}\else \'{c}\fi{}, and O. Vafek for helpful discussions.  This work was supported by the Department of Energy, contract DE-FG02-05ER46248 and the National High Magnetic Field Laboratory, the National Science Foundation, and the State of Florida.
	
The authors declare no competing financial interests

\section{Author Contributions}

Experiments were carried out by A.M.M., S.O., S.M., W.P.H., A.P.R. and P.L.K.  Samples were provided by K.F., M.I., and S.U.

%\bibstyle{natbib.sty}
\bibliography{Spin_Density_Modulation_Report}

\section{Supporting Online Materials}

\section{Materials and Methods}

	Our Bi$_2$Sr$_2$CaCu$_2$O$_{8+\delta}$ sample was post-processed by isotope exchange in 1 bar of flowing $^{17}$O ($\approx$ 90\% enriched) at 600 $^{\circ}$C for 48 h followed by annealing for 150 h at 450 $^{\circ}$C resulting in a superconducting transition temperature $T_c = 82$ K. Spatially resolved $T_1$ and $T_2$ measurements were taken by  Fourier transforming progressive saturation and Hahn echo sequences, respectively, with 16-pulse, phase alternation.  Spectra from the central transition of the planar oxygen O(1) were acquired for a wide range of relaxation recovery times and then the frequency domain of each one was subdivided into 32 intervals.  The recovery profiles in each interval were fit to a progressive saturation recovery\cite{mit01a} for  $T_1$ and an exponential for $T_2$.   Experiments presented here were performed at the National High Magnetic Field Lab in Tallahassee, Florida and at Northwestern University for fields from 4 to 30 T at a temperature of 4.2 K. The spectra were taken using a frequency sweep technique.  Before taking data at a particular applied magnetic field, we heated the sample above $T_c$ for 10 min followed by quench cooling to 4.2 K.  This procedure consistently produced reproducibly narrow spectra through many field cycles for which one indication is the systematic dependence of the linewidth on magnetic field, Fig.~2.  Data for some fields were also acquired at different temperatures, but are not reported here.  Measurements were also taken at magnetic fields below those shown in Fig. 2 and are reported separately \cite{mou10}.
	
	Spectra from the oxygen sites,\cite{che07} O(1) in the CuO$_2$ plane and O(2) in the SrO plane, are well-resolved at $T > 40$ K.  At lower temperatures the central transitions overlap.  Discrimination between O(1) and O(2) can be achieved by saturation of the O(2) resonance which is only weakly coupled to the electronic excitations in the CuO$_2$ plane and has an order of magnitude smaller $T_1^{-1}$ compared to O(1).  The relative intensities of the central and quadrupolar satellites in the O(1) spectrum, which do not overlap with the O(2) spectrum, are monitored down to 4.2 K to assure that the O(1) central transition has no contribution from O(2).  

\section{NMR linewidth}

The temperature of the transition from vortex liquid to vortex solid has been determined as a function of magnetic field by Chen {\it et al.}\cite{che07} in Bi2212, signaled by an abrupt increase in linewidth with decreasing temperature.  In the liquid state the time scale for vortex dynamics is much shorter than that of the NMR measurement, thereby averaging the local magnetic fields associated with vortices.  The experiments reported here are performed in the vortex solid phase.\\
\begin{figure}[b]
	\centering
	\includegraphics{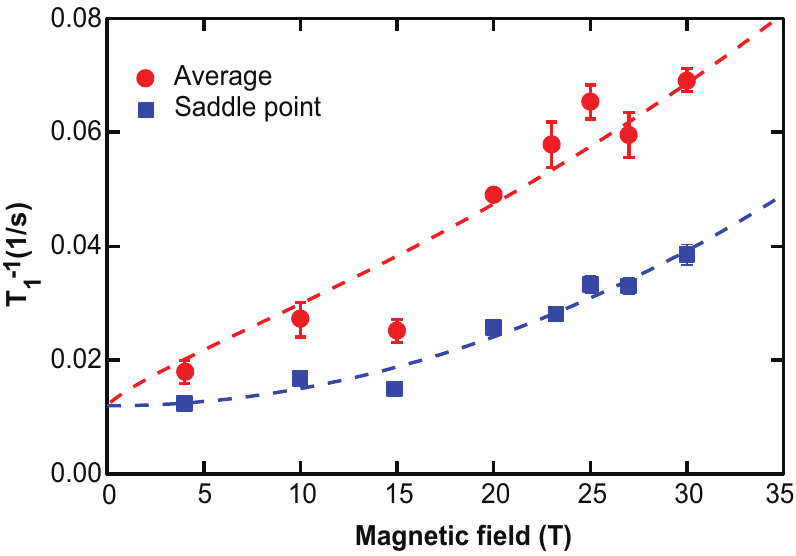}
	\caption{The average SLR rate, $T_1^{-1}$, as a function of magnetic field. The average over the vortex unit cell (red points) was obtained by weighting the measured  $T_1^{-1}$ by the spectrum, {\it i.e.} the probability distribution of local fields.  We expect that the contribution to the average $T_1^{-1}$ from the Volovik effect\cite{vol93} is proportional to H\,$log$H which, when added to the Zeeman contribution (green points, measured at the saddle point peak in the spectrum), fits the total average field dependence as shown by the dashed curve.  The intercept at $H=0$ is the thermal contribution to $T_1^{-1}$.}
	\label{AveT1}
\end{figure}
\indent It is instructive to compare the linewidth for $^{17}$O NMR reported here with that measured in YBCO in the normal state for various cations (Zn, Ni, and Li) substituted for copper in the CuO$_2$ plane generating an impurity local moment that contributes to the lineshape.  The  $^{17}$O NMR broadening we observe in Bi2212 is similar in magnitude to that reported for Zn in YBCO at the same field.  For this comparison we make an adjustment for the vortex concentration (Bi2212) to be the same as the cation impurity concentration (YBCO).  We assume the same hyperfine coupling for Bi2212 and infer, based on the NMR spectrum alone, that the vortex core could be modeled as a {\it local}  impurity with moment of $\sim 0.6 \mu_B$.   By local impurity we mean a perturbation that induces local field, RKKY-type oscillations with period $\sim \pi/k_F$,  which is about a factor of four  smaller than the modulation period we use here.  Although our analysis does not allow the numerical freedom to independently choose the period of the modulation for all applied fields, nonetheless, at the highest magnetic fields we can perform a $\chi^2$ analysis to simultaneously fit the measured  SLR rate and the spectrum in terms of the model of Eq.~1.  We find that the oscillation period is then constrained to be $30 \pm  5$ \AA,  as might be expected from the reports from STM and elastic neutron scattering.\cite{hof02,lak02}

\section{NMR relaxation}

\indent Spatially resolved NMR relaxation experiments have been carried out on YBCO aligned powders in previous work.\cite{cur00,mit01a,mit01b} However, the present single crystal samples of BSCCO show significantly higher quality results, in part because the quadrupolar satellites are fully resolved.  Nonetheless, both the YBCO and Bi2212 measurements  at low temperatures, $T < 15$ K,   and at high magnetic fields, $10 < H \leq 30$ T,  consistently indicate\cite{mit01a,mit01b} that at {\it each frequency position in the spectrum}, $T_{1}T$ is a constant and that $T_1^{-1}$ increases with applied magnetic field uniformly across the spectrum, except in the vortex core region.  This quadratic field dependence has been  attributed to the Zeeman shift in the nodal quasiparticle (NQ) energy.\cite{mit01a,mit01b}   The mechanism for relaxation with nodal quasiparticles includes thermal, Zeeman, and Doppler contributions. Based on the above observations, at fixed applied magnetic field, the dependence of the SLR rate on position in the vortex unit cell can be attributed to the Doppler shift in the NQ energy, ${\bf v}_F \centerdot {\bf p}_s$.  Since the SLR rate depends on the square of the local density of states (LDOS) and the  LDOS is  linear in the NQ energy, then the Doppler contribution to the SLR rate is $\propto p_s^2$ as displayed in Fig.~1(a). These arguments are presented in more detail by Mitrovic {\it et al.}\cite{mit01a}  \\
\indent The time derivative of the measured relaxation profile in the limit of short times, averaged over the entire spectrum, gives the average SLR rate.  This is proportional to the average of the square of the LDOS over the vortex unit cell whose spatial area is itself proportional to applied magnetic field.  Volovik\cite{vol93} pointed out that such an average is dominated by the Doppler term in the electronic excitation energy.  In Fig.~5 we show the magnetic field dependence of the average SLR rate, red points, together with the Zeeman contribution to the rate, green points,  measured at the peak in the spectrum, {\it i.e.} the saddle point. The difference between the data sets is due to the Volovik effect following the expected $H$\,log$H$ behavior which we show, together with the Zeeman contribution, as the (red) dashed curve in the figure.  This agreement confirms that the Doppler effect on NQ is a principal mechanism for relaxation and depends on position in the vortex unit cell.  Consequently, a measurement of $T_1^{-1}$ can distinguish between atoms in the NMR signal that are at different positions relative to the vortex core thereby providing spatial resolution within the CuO$_2$ plane.

\end{document}